\documentclass[prd,aps,twocolumn]{revtex4}
\usepackage{amsmath}
\usepackage{amssymb}
\usepackage{eucal}
\usepackage[all]{xy}
\input xypic
\numberwithin{equation}{section}
\begin{document}
\title{Novel Geometrical Models of Relativistic Stars.\\
      I. The General Scheme}
\author{P.~P.~Fiziev\footnote{ E-mail:\,\, fiziev@phys.uni-sofia.bg} }
\affiliation{Department of Theoretical Physics, Faculty of
Physics, Sofia University, 5 James Bourchier Boulevard,
Sofia~1164, Bulgaria.\\and\\ Joint Institute of Nuclear Research,
Dubna, Russia. }
\begin{abstract}
In a series of articles we describe a novel class of geometrical
models of relativistic stars. Our approach to the static
spherically symmetric solutions of Einstein equations is based on
a careful physical analysis of radial gauge conditions. It brings
us to a two parameter family of relativistic stars without stiff
functional dependence between the stelar radius and stelar mass.
It turns out that within this family there do exist relativistic
stars with arbitrary large mass, which are to have arbitrary small
radius and arbitrary small luminosity. In addition, point particle
idealization, as a limiting case of bodies with finite dimension,
becomes possible in GR, much like in Newton gravity.

\noindent{PACS number(s): 04.20.Cv, 04.20.Jb, 04.20.Dw}
\end{abstract}
%
\sloppy
\newcommand{\lfrac}[2]{{#1}/{#2}}
\newcommand{\sfrac}[2]{{\small \hbox{${\frac {#1} {#2}}$}}}
\newcommand{\ben}{\begin{eqnarray}}
\newcommand{\een}{\end{eqnarray}}
\newcommand{\la}{\label}
\maketitle
%
\section{Introduction}
Today the theory of relativistic stars is a well developed branch
of relativistic physics with many observational confirmations. It
is based on general relativity (GR), as a relativistic theory of
gravity, on quantum statistical physics, as a theory of
many-particle systems, and on the latest achievements of the
standard model (SM), as a modern theory of matter constituents.
(See the books \cite{books} and the large amount of references
therein.)

The role of quantum statistics of the Fermi gas in theory of
neutron stars has been clarified in the pioneering articles by
Chandrasekhar and Landau \cite{CL} on the non-relativistic ground
of Newton gravity. For this purpose was used the analogy with the
theory of white dwarfs \cite{Chandra}.

The beginning of the relativistic stelar theory can be found in
the pioneering articles by Schwarzschild \cite{Schwarzschild},
Tolmen, and Openheimer$\&$Volkov (TOV) \cite{TOV}, an important
further developments -- in \cite{developments}.

After the appearance of these articles the relativistic theory of
gravity of spherically symmetric static stars is widely accepted
as a well established issue. The further developments are related
with various considerations of the physics of stelar matter and
with a search of more realistic equations of state (EOS) of this
matter for different types of stars. This line of investigation is
continuously followed up to now, see \cite{EOS} and the references
therein.

In spite of general success of the relativistic theory of stars,
at present it can not be considered as a complete established
scientific area in its final form. (See, for example, the recent
review articles \cite{ns} and the references therein.)

Some difficulties in the explanation of the properties of very
dense stars, like neutron stars or eventual quark ones, are
suspected \cite{nsDiff} and need a proper explanation. The
currently used approach of continuous modification of EOS already
brought us to EOS, which are extension of known physical laws onto
domain where we have no experimental information. Therefore some
additional, more or less arbitrary assumptions are needed.

To some extend, a similar situation we may observe in the theory
of white dwarfs, see, for example, \cite{WD} and the references
therein, as well as some additional comments in Section V, B.

Unfortunately, we still do not have a complete set of
observational data for a direct confrontation of the present-days
relativistic theory with astrophysical observations. In particular
we do not have precise observational data both for the mass and
the radii of a given neutron star, we do not know the precise
upper limit of neutron star masses, e.t.c.

The existence of  such universal upper limit is a basic prediction
of the modern relativistic theory of stars, but the theory is not
able to give a definite predictions for the corresponding value,
due to the uncertainties in EOS. As a result, all observed massive
compact dark objects with {\em gravitational} mass $m_*\gtrsim
(5-10)\times m_\odot$ are automatically interpreted as a
candidates for black holes, despite of the fact that there still
do not exist undisputable {\em direct} observational evidences for
existence of such exotic objects with their non-avoidable
attribute - the event horizon \cite{BH}.

The fast development of this scientific domain calls for a further
investigation of different aspects of the general theory,
including its basic assumptions.

In the present series of articles we will not consider the EOS
problem, nor the complicated dynamical problems of rotating stars,
or stelar oscillations. Here we will reconsider some basic
features of relativistic theory of gravity when applied to the
study of stelar physics in the simplest static spherically
symmetric case. We shall show that there exist new classes of
models for relativistic stars, thus enlarging essentially the
general theoretical scheme. We hope that the new relativistic
models may lead to a better understanding of the real
astrophysical observations.

Here we utilize a new approach to the spherically symmetric static
solutions to Einstein equations (EE), based on careful analysis of
the radial gauge. Recently this approach brought into the world a
new two-parameter family of solutions of EE with a massive point
source and some unexpected physical consequences, see \cite{F03}.
One of them is that in the point particle problem the global
analytical properties of the solutions of EE in complex plain of
the radial variable are fixing this variable in a unique way,
together with the corresponding boundary conditions. Similar
phenomenon is well known in the theory of analytical functions:
they are unambiguously defined by their singular points in complex
domain.

In mathematical sense this way were derived the {\em fundamental}
static spherically symmetric solutions of EE. These solutions are
analogous to the fundamental solutions of classical Poison
equation with point source. Here we are extending this approach to
the theory of relativistic stars.

\section{The Physical Consequences of the Choice of Radial Gauge
         in the Stelar Problem}

The EE determine the solution of a given physical problem up to
four arbitrary functions, i.e., up to a choice of coordinates.
This reflects the well known fact that GR is a gauge theory.
According to the standard textbooks \cite{books}, the fixing of
the gauge in GR in a {\em holonomic} frame is represented by a
proper choice of the quantities
\ben\bar\Gamma_\mu\!=\!-{{1}\over{\sqrt{|g|}}}g_{\mu\nu}
\partial_\lambda\left(\sqrt{|g|}g^{\lambda\nu}\right),\la{Gammas}\een which
emerge when one expresses the 4D d'Alembert operator in the form
$g^{\mu\nu}\nabla_\mu\nabla_\nu=g^{\mu\nu}
\left(\partial_\mu\partial_\nu-\bar\Gamma_\mu\partial_\nu\right)$.
Unfortunately, up to now physically reasonable principles for the
choice of the gauge in GR are not known. Moreover,  at present
many of the relativists are thinking that this is not a physically
essential GR problem.

We shall call the change of the gauge fixing expressions
(\ref{Gammas}), {\em without} any preliminary conditions on the
analytical behavior of the used functions, a gauge transformations
in a {\em broad} sense. This way we essentially expand the class
of the gauge transformations, we intend to discuss, looking for a
physically meaningful choice of the gauge conditions
(\ref{Gammas}).

In the static spherically symmetric problems the structure of the
space-time is
$\mathbb{M}^{(1,3)}=T_t(1)\times\mathbb{M}^{(1)}_r\times SO(3)$.
There exists unambiguous choice of the global time $t$ on the 1D
time-translations group $T_t(1)$ and of the angle variables
$\theta$, $\phi$ -- on the $SO(3)$ group space. These variables
are unambiguously fixed by symmetry reasons. In proper units (in
which the velocity of light is $c=1$) this choice yields the
familiar form of the space-time interval: \ben
ds^2=g_{tt}(r)\,dt^2\! +g_{rr}(r)\,dr^2\!-\!\rho(r)^2(d\theta^2\!+
\!\sin^2\theta\,d\phi^2)\hskip .3truecm \la{ds0}\een with unknown
functions $g_{tt}(r)>0,\,g_{rr}(r)<0,\,\rho(r)$.

Thus the form of three of the gauge fixing coefficients
(\ref{Gammas}):\, $\bar\Gamma_t\!=\!0,\,\,\bar\Gamma_\theta\!=
-\!\cot\theta,\,\,\bar\Gamma_\phi\!=\!0$ is fixed by symmetry
reasons, but the quantity
\ben\bar\Gamma_r\!=\left(\!
\ln\left({\sqrt{-g_{rr}}\over{\sqrt{g_{tt}}\,\rho^2}}
\right)\!\right)^\prime\!=\! \left(\!
\ln\left({\rho^\prime{\sqrt{-g_{\rho\rho}}}\over
{\rho^2\,\sqrt{g_{tt}}}}\right)\!\right)^\prime, \la{Gamma_r}\een
and, equivalently, the function $\rho(r)$ are still not fixed.
Here and further on, the prime denotes differentiation with
respect to the variable $r$.

The physical and the geometrical meaning of the radial coordinate
$r$ is not defined by symmetry reasons and is unknown \textit{a
priori} \cite{Eddington, F03}. The only clear thing is that its
value $r=0$ corresponds to the center of $SO(3)$ symmetry. In the
case of relativistic stars with {\em regular} distribution of
matter this 3D-space point is the physical center of the star,
where the mass $m(r)$, surrounded by a sphere with {\em
coordinate} radius $r=0$, is $m(0)=0$.

We shall use this {\em physical} property of the mass as a
definition of the star's center $C$, because it does not depend on
the choice of the radial variable $r$. Thus the mass $m$ can be
used to find the geometrical place at which the proper radial
variable $r$ must equals zero. This is the main specific feature
of our approach to the theory of the relativistic stars.

In contrast to the radial variable $r$, the quantity $\rho$ has a
clear geometrical and physical meaning: $\rho$ defines the area
$A_\rho=4\pi\rho^2$ of a centered at the center $C$ sphere with
"area radius" $\rho$. From physical point of view one can refer to
this quantity as {\em "a luminosity variable``} (or "a luminosity
radius"), because the luminosity of distant physical objects is
reciprocal to $A_\rho$. In other words, the variable $\rho$
describes the spherically symmetric spreading of energy of any
kind.

We refer to the choice of the function $\rho(r)$ as a choice of
radial gauge in a broad sense \cite{F03}, allowing, in general,
singular changes of the variable $r$. We call the freedom of
choice of the function $\rho(r)$ "{\em a rho-gauge freedom}" in a
{\em broad} sense, and any definite choice of  function $\rho(r)$
-- "{\em a rho-gauge fixing}".

At first glance the fixing of the function $\rho(r)$ seems to be
rather arbitrary and without any physical significance.

From geometrical point of view the choice of the radial gauge
defines an imbedding of the 1D quotient space
$\mathbb{M}^{(1)}_r\!=
\!\left(\mathbb{M}^{(1,3)}/T_t(1)\right)\!/\!SO(3)$ into the
space-time $\mathbb{M}^{(1,3)}$.

For fixing of this {\em additional} mathematical structure one
needs some physical conditions like boundary conditions, or
conditions for fixing of the number and the character of singular
points of the solution of EE in the whole complex domain. This was
demonstrated in \cite{F03} for the case of fundamental singular
solutions of EE with massive point source. These additional
conditions play an essential role in the problem, because they are
determining the global analytical properties of the solutions.
Actually they define the very manifold $\mathbb{M}^{(1,3)}$.

The EE are holomorphic ones and their solutions must be studied in
the whole complex domain of corresponding variables. The very EE
do determine only the {\em local} structure of
$\mathbb{M}^{(1,3)}$. In our case the change of the function
$\rho(r)$ will be not a simple change of the labels of space-time
points, if it changes the additional conditions, which fix the
analytical properties of the manifold $\mathbb{M}^{(1,3)}$ in the
whole complex domain \cite{F03}.

Our present consideration illustrates this important juncture on
the more physical example of solar models:

It is obvious that physical results of any theory must not depend
on the choice of the variables. In particular, these results must
be invariant under changes of coordinates. This requirement is a
basic principle not only in GR. It is fulfilled for any {\em
already fixed} mathematical problem.

Nevertheless, the change of the interpretation of the variables
may change the very mathematical problem and some physical
results, because we are using the variables according to their
interpretation. For example, if we are considering the luminosity
variable $\rho$ as a radial variable of the problem, it seems
natural to put the center of the star at the point $\rho=0$. In
general, we may obtain a physically different stelar model, if we
are considering another variable $r$ as a radial one: in this case
we shall place the star's center at a different geometrical point
$r=0$, which now seems to be the natural position of the physical
center $C$. The relation between these two geometrical "points"
and between the corresponding stelar models strongly depends on
the choice of the function $\rho(r)$, i.e. on the radial gauge.

Thus, applying the same physical requirements, like
\ben m|_{\hbox{\small at the center}\,\, C}=0,\la{m_cond}\een
in different "natural" variables, we arrive at different physical
models, because we are solving EE under different boundary
conditions. One has to find a theoretical or an experimental
reasons to resolve this essential ambiguity, or one has to accept
it as an non-avoidable component of the theory, recovering its
physical meaning and its proper usage.

The physical center $C$ of the star is placed at the point $r=0$
by definition. To what value of the luminosity variable
$\rho_C=\rho(0)$ corresponds the {\em real} position of the center
$C$ is not known {\em a priori}. This depends strongly on the
choice of the rho-gauge function $\rho(r)$. One can not exclude
such nonstandard behavior of the physically {\em reasonable} gauge
function $\rho(r)$, which leads to some value $\rho_C > 0$
\cite{F03}.

This very interesting novel possibility emerges in curved
space-times due to their unusual geometrical properties and is not
supported by our Euclidean experience. It will be the main subject
of study in the present series of articles. Such possibility was
discovered at first in the original pioneering article by
Schwarzschild \cite{Schwarzschild1} and discussed by Brillouin
\cite{Brillouin}, but at present it is widely ignored. A physical
necessity of considering values of the $\rho$-variable, not less
the Schwarzschild radius, was stressed by Dirac \cite{Dirac}, too.

The present-days standard theory of relativistic stars is based on
the Hilbert radial gauge (HG): $\rho(r)\equiv r$. In this gauge
the center of the star is placed at the point $\rho_C=\rho(0)=0$.
This rather arbitrary additional condition was at first utilized
by Schwarzschild in his simple model of incompressible stars
\cite{Schwarzschild}. There he had used formal mathematical
reasons to be able to fix this way one of the integration
constants. Actually he had postulated the {\em global} geometrical
properties of the stelar center $C$ in curved space-time, adopting
the ones, which take place in the non-relativistic Euclidean case.

The {\em local} reason seems to be Eq. (\ref{m_cond}), which
entails asymptotically flat 3D metric in a small enough vicinity
of the stelar center $C$. According to GR, the spherically
symmetric distribution of the masses outside this vicinity does
not influence the flat geometry around $C$.

Nevertheless, one has to take into account that there is no
guaranty that starting from some luminosity $L_* \rightleftarrows
\rho_*$ at the stelar surface, and going trough the {\em curved}
3D space back to the center $C$, defined by Eq. (\ref{m_cond}), we
will reach the value $\rho_C=0$. This is what we mean by "global"
property of center $C$. This property of the center $C$ depends on
properties of the interior solution of stelar problem in {\em the
whole} interior domain. From point of view of TOV system of
differential equations, such property of the center $C$ depends on
the {\em global} properties of the inner solution.

The assumption $\rho_C=0$ has a strong influence on the further
development of the theory of relativistic stars. In particular, it
forces one to impose a regularity condition at the point $\rho=0$
both on the matter distribution and on the solutions of EE. As a
result, one uses {\em only} a very specific solutions of TOV
equations for relativistic stars \cite{books}--\cite{ns}. These
solutions form a set of zero measure in the variety of all
solutions of the problem. Thus one is forced to ignore the vast
majority of the solutions, which are of general type and do not
obey the regularity condition {\em at the point} $\rho=0$.

The novel solutions of EE for massive point particle, discovered
in \cite{F03}, raise a new understanding of the role of the
luminosity variable $\rho$. Here we shall show that in the
relativistic theory of stars the same approach allows a
consideration of {\em all} solutions of the TOV equations in a
physically meaningful way. This way we essentially enrich the
relativistic theory of stars.

An extremely important consequence of our more wide treatment of
stelar models is the existence of relativistic stars with {\em
arbitrary large mass}, and, at the same time, with arbitrary small
geometric radius and arbitrary small luminosity. This unexpected
possibility will be mathematically proved for incompressible
relativistic stars in a subsequent article.

The strong nonlinearity both of the differential equations and the
boundary conditions may yield, in general, several different
classes of new solutions. This resembles the real situation,
illustrated by the well known Hertzsprung-Russell diagram for
stars in Nature \cite{HR}.

Because of presence of the additional parameter $\rho_C$ in the
general solutions of TOV equations, the total mass of the star
$m_*$ is not a function {\em only} of its coordinate radius $r_*$
(or geometrical radius $R_*$, or luminosity radius $\rho_*$) and
may vary independently of it. Due to this property, our approach
gives for the first time a possibility to consider the point
particles in GR as a limiting case of a body with finite
dimension, much like in the Newton theory of gravity. This
important new feature will be described in another subsequent
article.

The considerations in this first of series of articles has a
preliminary character, setting in a new way the relativistic
theory of gravity in the stelar physics. This article does not aim
a construction of a specific models of relativistic stars. Here we
give only the new general scheme for such considerations. Due to
technical reasons we describe specific models of relativistic
stars with different EOS and other further developments elsewhere.

\section{Solution of the  Extended TOV System in Hilbert Gauge}
The luminosity variable $\rho$ gives a very convenient description
of stelar structure in real domain. The use of this variable
ensures {\em a local} radial-gage-invariance of the approach to
solution of this problem. Therefore, keeping the traditions, we
shall work in this Section in HG. In our consideration the values
$\rho<\rho_C$ have no physical meaning. The values $\rho\in
[\rho_C,\rho_*]$ describe the inner domain and the values $\rho\in
(\rho_*,\infty)$ correspond to the exterior vacuum domain outside
the star.

Then the inner metric (\ref{ds0}) for a static spherically
symmetric star is defined by the metric components \ben
g_{tt}(\rho;\rho_*,\rho_C)\!=\!e^{2\varphi(\rho;\rho_*,\rho_C)}
,\nonumber \\
g_{\rho\rho}(\rho,;\rho_*,\rho_C)\!=\!{-1\over{1\!-\!2m(
\rho;\rho_*,\rho_C)/\rho}}. \la{sg}\een Here and further on we are
using units $c=G_N=1$.

The mass $m(\rho;\rho_*,\rho_C)$, surrounded by a sphere with a
luminosity radius $\rho$, obeys the first TOV equation: \ben
{{dm}\over{d\rho}}=4\pi\varepsilon\rho^2 > 0, \la{TOV1}\een
supplemented by the boundary conditions \ben
m(\rho_C;\rho_*,\rho_C)=0
,\,\,\,\,m(\rho_*;\rho_*,\rho_C)=m(\rho_*,\rho_C). \la{BCTOV1}\een
The first condition is the definition of the physical center $C$
of the star, placed at a position with {\em unknown} value of the
luminosity variable $\rho_C\geq 0$. The second one defines the
total mass of the star $m(\rho_*,\rho_C)$, obtained using the {\em
unknown} value of the luminosity variable $\rho_*\geq \rho_C$ at
the edge of the star.

As a consequence of Eq.(\ref{sg}) and (\ref{TOV1}) we obtain for
any admissible value of $\rho_C$ the relation
$-g_{\rho\rho}(\rho_C;\rho_*,\rho_C)=1$.

The pressure $p(\rho;\rho_*,\rho_C)$ obeys the equation
\ben {{dp}\over{d\rho}}=-{{(p+\varepsilon)(m+4\pi\rho^3 p)}
\over{\rho(\rho-2m)}} < 0, \la{TOV2} \een
where $\varepsilon(\rho;\rho_*,\rho_c)$ is the energy density. The
boundary conditions for this equation are:
\ben p(\rho_C;\rho_*,\rho_C)=p_C(\rho_*,\rho_C)
,\,\,\,\,p(\rho_*;\rho_*,\rho_C)=0. \la{BCTOV2}\een
Here $p_C$ is the pressure at the star center $C$. The second
condition defines the physical edge of the star.

One has to extend the above TOV system adding the equation for the
proper mass $m_0(\rho;\rho_*,\rho_c)$ of the star in the sphere
with luminosity radius $\rho$:
\ben
{{dm_0}\over{d\rho}}=4\pi\varepsilon\rho^2\sqrt{-g_{\rho\rho}} >
0, \la{TOV3}\een
together with the boundary conditions \ben
m_0(\rho_C;\rho_*,\rho_C)=0
,\,\,\,\,m_0(\rho_*;\rho_*,\rho_C)=m_0(\rho_*,\rho_C).\hskip
.5truecm \la{BCTOV3}\een and the equation for the gravitational
potential $\varphi(\rho;\rho_*,\rho_C)$: \ben
{{d\varphi}\over{d\rho}}={{m+4\pi\rho^3 p}
\over{\rho(\rho-2m)}}>0,\la{TOV4}\een together with the boundary
conditions \ben
\varphi(\rho_C;\!\rho_*,\rho_C)\!=\!\varphi_C(\rho_*,\rho_C)
,\,\varphi(\rho_*;\!\rho_*,\rho_C)\!=\!\varphi(\rho_*,\rho_C).\hskip
.5truecm \la{BCTOV4}\een

To have a closed system of mathematical equations one has to add
the EOS. It can be defined in different equivalent forms. The most
convenient for our general considerations is the following one:
\ben \varepsilon=\varepsilon(p). \la{EOS}\een

It is useful to introduce, too, the quantity
\ben w=p/\varepsilon.\la{w}\een

 As a result of our consideration we see that
the description of stelar structure leads to a correct
mathematical boundary problem with unknown ends $\rho_C$ and
$\rho_*$.

\section{Solution of a Cauchy Problem as a Method of Solution of
the Stellar Boundary Problem}

As we have seen in the previous Section, the stelar structure is
determined by solution of the boundary problem, described by Eq.
(\ref{TOV1})-(\ref{EOS}). It is remarkable that in the simple case
at hand the subsystem of differential equations (\ref{TOV1}),
(\ref{TOV2}) splits and can be solved independently of the other
equations in ETOV system. Using the first of the conditions
(\ref{BCTOV1}) and (\ref{BCTOV2}) as initial conditions for this
subsystem, one obtains the solutions of the corresponding Cauchy
problem in the form \cite{books}: \ben
m=m(\rho;\rho_C,p_C),\,\,\,\,p=p(\rho;\rho_C,p_C).\la{Cauchy}\een
Using the already known function $p(\rho;\rho_C,p_C)$, one can
solve the second of the Eq.(\ref{BCTOV2}), written in the form
$p(\rho_*;\rho_C,p_C)=0$ with respect to the quantity $p_C$.  Thus
one obtains the pressure at the center $C$ in the form
$p_C=p_C(\rho_*,\rho_C)$. Now, substituting this function in the
known expressions $m(\rho;\rho_C,p_C)$ and $p(\rho;\rho_C,p_C)$,
and solving the corresponding integrals, one obtains the solution
of the whole boundary problem, described in the previous Section,
in the form:
\begin{subequations}\label{SolBound:xyzt}
\ben
m(\rho;\rho_*,\rho_C)=m(\rho;\rho_C,p_C(\rho_*,\rho_C)),\hskip
1.25truecm \la{SolBound:x}
\\ p(\rho;\rho_*,\rho_C)=p(\rho;\rho_C,p_C(\rho_*,\rho_C)),
\hskip 1.5truecm \la{SolBound:y}\\
m_0(\rho;\rho_*,\rho_C)=4\pi\int_{\rho_c}^{\rho}
{{\varepsilon(\rho;\rho_*,\rho_C)\rho^2 d\rho
}\over{\sqrt{1-m(\rho;\rho_*,\rho_C)/\rho}}},
\la{SolBound:z}\\
\varphi(\rho;\rho_*,\rho_C)=\varphi_C(\rho_*,\rho_C)+
\hskip 2.5truecm\nonumber\\
\int_{\rho_c}^{\rho}{{m(\rho;\rho_*,\rho_C)+4\pi\rho^3
p(\rho;\rho_*,\rho_C) }\over{{\rho \big(
\rho-2m(\rho;\rho_*,\rho_C)\big)}}}d\rho.
\,\,\,\,\,\la{SolBound:t}\een
\end{subequations} Here \ben
\varepsilon(\rho;\rho_*,\rho_C)=\begin{cases}
\varepsilon\big(p(\rho;\rho_*,\rho_C)\big), & \text{if $\rho\in
[\rho_C,\rho_*]$}; \cr 0, & \text{if $\rho\in
(\rho_*,\infty)$}\,\,\,\,\,\la{epsilon}\end{cases}\een is obtained
making use of Eq.(\ref{EOS}).

The exterior solution in HG is well known: \ben
g_{tt}(\rho;\rho_*,\rho_C)\!=\!{{-1}\over
{g_{\rho\rho}(\rho;\rho_*,\rho_C)}}\!=
\!1-{{2m(\rho_*,\rho_C)}\over{\rho}}.\,\,\,\,\, \la{exterior}\een
It gives an interpretation of the constant $m(\rho_*,\rho_C)$ as a
Keplerian mass of the star.

The inner solution (\ref{SolBound:xyzt}) depends on the function
$\varphi_C(\rho_*,\rho_C)$ which can be determined using Birkhoff
theorem. The gravitational field at the edge of the star depends
only on the total mass $m(\rho_*,\rho_C)$ and matches the exterior
vacuum solution, which is unique, up to choice of radial gauge.
When written in HG, the matching condition gives
$\varphi(\rho_*,\rho_C)=\ln
g_{tt\,*}=\ln\sqrt{1-2m(\rho_*,\rho_C)/\rho_*}$. Then from the
last equation  (\ref{SolBound:t}) one obtains
\ben \varphi_C(\rho_*,\rho_C)=\ln\sqrt{1-2m(\rho_*,\rho_C)/\rho_*}-\nonumber \\
\int_{\rho_c}^{\rho_*}{{m(\rho;\rho_*,\rho_C)+4\pi\rho^3
p(\rho;\rho_*,\rho_C) }\over {\rho \big(
\rho-2m(\rho;\rho_*,\rho_C)\big)}}d\rho. \la{phiC}\een As a result
\ben \varphi(\rho;\rho_*,\rho_C)=\ln\sqrt{1-2m(\rho_*,\rho_C)/\rho_*}-\nonumber\\
\int_{\rho}^{\rho_*}
{{m(\rho;\rho_*,\rho_C)+4\pi\rho^3p(\rho;\rho_*,\rho_C)} \over
{\rho \big( \rho-2m(\rho;\rho_*,\rho_C)\big)}}d\rho
\la{varphi}\een and the inner solution depends only on the two
parameters $\rho_C$ and $\rho_*$ with  unknown values.

The geometrical radius of the star
$R_*=R(\rho_*,\rho_C)=\int_{\rho_C}^{\rho_*}\sqrt{-g_{\rho\rho}}d\rho=
\int_0^{r_*}\sqrt{-g_{rr}}dr$ is \ben
R(\rho_*,\rho_C)=\int_{\rho_c}^{\rho_*}{\rho\,{d\rho}
\over{\sqrt{\rho\left(\rho-2m(\rho;\rho_*,\rho_C)\right)}}}.
\la{R}\een

Thus for a given EOS we arrived at a two-parameter family of
relativistic stars.  Up to now the solutions have been
parameterized by luminosity variables $\rho_C$ and $\rho_*$.

A similar procedure, based on solution of back Cauchy problem with
initial point at the edge of the star, $\rho_*$, illustrates in
the best way our {\em physical} definition of the center of star
$C$:

We can solve the subsystem of differential equations (\ref{TOV1}),
(\ref{TOV2}) under much more physical initial conditions -- fixing
in arbitrary way the directly measurable mass $m_*>0$ and
luminosity radius $\rho_*>\rho_G=2 m_*$ and using the value of
pressure $p_*=0$ at the edge of the star. Now we can integrate the
differential equations back with respect to the variable $\rho$.
According to Eq. (\ref{m_cond}), the center $C$ of the star is
defined as a point $\rho_C<\rho_*$, at which
$m(\rho_C,\rho_*,\rho_C)=0$. Finding this way
$\rho_C=\rho_C(m_*,\rho_*)$, we obtain the not-directly-measurable
pressure
$p_C\left(=p_C(\rho_*,\rho_C)=p(\rho_C,\rho_*,\rho_C)\right)$ at
the center $C$, which, itself, is hidden for us, from
observational point of view.

Obviously, the widespread in the literature \cite{books} stiff
relation $m_*=m_*(\rho_*)$ will appear {\em only} if we pose by
hands the commonly adopted extra condition $\rho_C(m_*,\rho_*)=0$,
although there are no physical reasons to do this.

It is clear, that the procedure, based on backward integration,
lies on much more physical ground, than the standard one. It is
complete equivalent to the traditional procedure, if we use {\em
unknown} value of the luminosity variable $\rho_C$  for solution
of Cauchy problem with the center of the star $C$ as starting
point.

\section{The Mappings $Rm$ and $Rm_0$}
\subsection{The Mapping $Rm$} If one solves the algebraic
equations \ben
R(\rho_*,\rho_C)=R_*,\,\,\,\,\,m(\rho_*,\rho_C)=m_*\la{m*R*}\een
with respect to the variables $\rho_*$ and $\rho_C$, expressing
them as a functions of the variables $R_*$ and $m_*$, one obtains
a complicated nonlinear mapping \ben \{R_*,m_*\} \xymatrix@1{ {}
\ar[r]^{Rm} & {}} \{\rho_*(R_*,m_*),\rho_C(R_*,m_*)\}.
\la{map}\een The study of this mapping is a basic physical problem
in our approach to the relativistic stelar structure.

This way one can parameterize the solutions of ETOV equations for
relativistic stars with a fixed EOS by the two parameters $R_*$
and $m_*$, which are {\em directly} measurable.

As we see, in our model of relativistic stars of most general
type, the theory of gravity in HG does not yield a functional
dependence between the mass $m_*$ and the radius $R_*$, only. For
a fixed value of $R_*$, according to Eqs. (\ref{m*R*}),
(\ref{map}), the mass of the star $m_*$ can still vary.

To obtain a stiff functional dependence between the mass $m_*$ and
the radius $R_*$ one must introduce some auxiliary condition. In
the commonly accepted models of relativistic stars the role of
such condition plays the assumption $\rho_C=0$, which seems to be
not necessary from physical point of view. If imposed, this extra
condition, together with relations (\ref{map}), gives the well
known stiff functional relation $m_*=m_*(R_*)$.

\subsection{The Mapping $Rm_0$} If one solves the algebraic
equations \ben
R(\rho_*,\rho_C)=R_*,\,\,\,\,\,m_0(\rho_*,\rho_C)=m_{0*}\la{m0*R*}\een
with respect to the variables $\rho_*$ and $\rho_C$, expressing
them as a functions of the variables $R_*$ and $m_{0*}$, one
obtains another complicated nonlinear mapping \ben \{R_*,m_{0*}\}
\xymatrix@1{ {} \ar[r]^{Rm_0} & {}}
\{\rho_*(R_*,m_{0*}),\rho_C(R_*,m_{0*})\}. \la{map0}\een  This way
one can parameterize the solutions of ETOV equations for
relativistic stars with a fixed EOS by the two basic parameters
$R_*$ and $m_{0*}$.

The study of the mapping $Rm_0$ (\ref{map0}) is the second basic
physical problem in our approach to the relativistic stelar
structure in HG.

In contrast to the Keplerian mass $m_*$, which depends on the
concentration of the fixed amount of matter in a given star, its
proper mass $m_{0*}$ is independent of this concentration and
characterizes the very amount of matter, the star is build of.

If imposed, the extra condition $\rho_C\!=\!0$ now yields a stiff
functional dependence $R_*\!=\!R_*(m_{0*})$ for any given EOS.
Combined with the relation $m_*\!=\!m_*(R_*)$ from the previous
Subsection, it leads to another stiff dependence:
$m_*\!=\!m_*(m_{0*})$.

The above stiff relations are the most important specific
prediction of the relativistic theory of gravity {\em with}
auxiliary condition $\rho_C\!=\!0$.

We have to stress that in Newton theory of gravity, which is known
to describe well enough the physics of large class of real stars
\cite{books}, including the Sun \cite{Bahcall}, as well as some
features of white dwarfs \cite{Chandra}, \cite{WD}, we have
analogous stiff functional relations, with similar origin -- the
regularity condition at the Euclidean point $r_C=0$.

At the same time the functional dependance between corresponding
quantities in the Newtonian theory is essentially different, in
comparison with the standard relativistic theory. In particular:
a) There we do not have two different masses $m_*$ and $m_{0_*}$,
because $m_*\equiv m_{0_*}$; b) In Newtonian models of spherically
symmetric bodies with different EOS, as a role, we have no
limitations on the mass $m_*$ and the radius $R_*$, due to the
requirement to have a regular solutions at the stelar center $C$.
The restrictions on the mass $m_*$ appear, as an exception, only
in the limit of degenerate {\em ultra-relativistic} matter
\cite{Chandra}, which is not a realistic case.

This is in a sharp contrast to the GR theory of stars, based on
the condition $\rho_C=0$, in which we have restrictions on the
mass $m_*$ for any EOS \cite{books}, \cite{Chandra},
\cite{Schwarzschild}, \cite{polytrops}.

In our more general geometrical models of GR stars the regularity
condition at center $C$ with $r_C=0$ are satisfied, because
$\rho_C=\rho(0)\neq 0$. Here we do not have stiff relations of the
discussed type, without imposing some additional extra conditions.

To check the existence of stiff relations between $m_*$, $m_{0*}$
and $R_*$ in Nature, one has to analyze properly the observational
data. The absence of stiff relations would lead to a dispersion of
the observational data in a large domain of the corresponding
variables. In the opposite case -- if the stiff relations take
place in physical reality, the observational data have to show a
clear functional dependence between the corresponding quantities
for some class of real stars with fixed EOS and matter content,
which are in the same instant state.

This phenomenon can help us to test the validity of relativistic
theory of gravity with $\rho_C=0$ (or with some other extra
condition) in stelar physics, performing a precise analysis of the
observational data.

Even a cursory look at the Hertzsprung-Russell diagram \cite{HR}
will convince us that the observations may not support the
standard relativistic theory of stars in HG with the extra
condition $\rho_C=0$ adopted:

In the Hertzsprung-Russell diagram we see a big dispersion of the
temperature-luminosity positions of stars with different masses
$m_*$ and radii $R_*$. Unfortunately it is not clear whether the
(non)existence of stiff relations can be mask completely by the
strong dependence on EOS and instant time state of the star, which
changes essentially during the time evolution.  On the
Hertzsprung-Russell diagram we are witnessing some mixture of
different effects, due to too many physical factors. This is a
serious obstacle for making some definite conclusions about the
problem, we are discussing.

Without any doubts, the best candidates for such analysis are the
white dwarfs. Their EOS is well known and fixed. In addition, some
observational information for their radiuses and masses is
available \footnote{The author is grateful to Eva-Maria Pauli and
to M. Miller for discussions on white dwarfs' physics and for
information about available data. Special tanks to Eva-Maria Pauli
for sending the file of her PhD thesis.}.

The corresponding stiff mass-radii relations were established on
the basis of theory of degenerate stelar matter and studied in
details in \cite{Chandra}. According to Provencal et al. (1998)
\cite{WD}, "One might assume that a theory as basic as stellar
degeneracy rest on solid observational ground, yet this is not the
case. Comparison between observation and theory has shown
disturbing discrepancies ...".

Actually, a relatively large dispersion of observational data for
masses and radii of white dwarfs are observed \cite{WD}. Its
explanation, on the basis of standard relativistic theory of
stars, forces one to accept a doubtful variations of the matter
content of the white dwarfs. For example, a possible explanation
of too small radii of some white dwarfs is the assumption about
the existence of iron-reach core in them. According to Panei et
al. (2000) \cite{WD}: "Obviously, such result is in strong
contradiction with the standard predictions of stelar evolutionary
calculations, which allow for an iron-rich interior only in the
case of presupernova objects".

Taking into account this situation, it seems interesting to
analyze the existing data for white dwarfs mass-radii relation
from point of view of the two parameter family of novel
geometrical models of relativistic stars, presented here. We
intend to perform such analysis elsewhere.

\subsection{The 2D Domains $\mathbb{D}^{(2)}_{\rho_*,\rho_C}$,
$\mathbb{D}^{(2)}_{R_*,m_*}$ and $\mathbb{D}^{(2)}_{R_*,m_{0*}}$}

In our new model of relativistic stars the parameters
$\{\rho_*,\rho_C\}\in \mathbb{D}^{(2)}_{\rho_*,\rho_C}\subseteq
\mathbb{R}^{(2)}$  vary in some 2D domain
$\mathbb{D}^{(2)}_{\rho_*,\rho_C}$, restricted by the conditions:
\begin{subequations}\label{Drhorho:xyz}
\ben
\renewcommand{\theequation}{\theparentequation\alpha{equation}}
 0\leq\rho_C\leq\rho_*, \la{Drhorho:x}\\
0<2m(\rho_*,\rho_C)<\rho_*, \la{Drhorho:y}\\
0<p_C(\rho_*,\rho_C)<\infty.\la{Drhorho:z}\een
\end{subequations}

These conditions determine the physical 2D domains
$\mathbb{D}^{(2)}_{R_*,m_*}$ and $\mathbb{D}^{(2)}_{R_*,m_{0*}}$
of the stelar parameters $R_*>0$, $m_*\geq 0$ and $m_{0*}\geq 0$
in the mappings (\ref{map}) and (\ref{map0}).

The form of the domains $\mathbb{D}^{(2)}_{\rho_*,\rho_C}$,
$\mathbb{D}^{(2)}_{R_*,m_*}$ and $\mathbb{D}^{(2)}_{R_*,m_{0*}}$
depends on the EOS. In general, its determination is a hard
theoretical problem. Its solution is important for observational
tests of the existence of the stiff relations, described in the
previous two Subsections.

\section{Scale Properties of the ETOV and HG Scale Invariant
Quantities and Relations}

An important general property of the equations
(\ref{TOV1})--(\ref{TOV4}) was discovered by Bondi in 1964. This
is their formal invariance under the scaling transformations with
a constant coefficient $\lambda$ \cite{developments}: \ben
\rho\!\to\!\lambda\rho,\,m\!\to\!\lambda m,\,m_0\!\to\!\lambda m_0
,\,\varepsilon\!\to\!\lambda^{-2}\varepsilon,\,p\!\to\!\lambda^{-2}
p.\hskip .6truecm\la{lambda}\een (See, too, the articles by
Hartle, by Ellis et al., and by Collins in \cite{developments}.)

It is obvious that the quantities $g_{tt}$, $\varphi$,
$g_{\rho\rho}$  and $w$ are $\lambda$-invariant.

If, and only if, $w=const$, the EOS (\ref{EOS}) is
$\lambda$-invariant and the solutions of the whole ETOV system
will have a self-similar behavior, see  the articles by Collins
and by Rendal\&Schmidt in \cite{developments}.

Instead of the local binding energy $\Delta
m(\rho,\rho_*,\rho_C):=m_0-m$, which is not $\lambda$ -invariant,
one can consider the ratio
\ben\varrho(\rho,\rho_*,\rho_C):=m/m_0\in (0,1).\la{varrho}\een

It measures in a $\lambda$ -invariant way the local mass defect of
the star mater, i.e. the mass defect in the sphere with luminosity
radius $\rho$ and center $C$.

Another important $\lambda$-invariant local (in the above sense)
quantity is
$f(\rho,\rho_*,\rho_C)=\varrho(\rho,\rho_*,
\rho_C)^2-g_{tt}(\rho,\rho_*,\rho_C)+1$.
In the case at hand it has the form
\ben f(\rho,\rho_*,\rho_C)=\left({{m}\over{m_0}}\right)^2+{{2
m}\over{\rho}}=\varrho^2+\varsigma^2,\la{Frho}\een
where $\varsigma^2\!=\!{{2 m}\over{\rho}}\geq 0$ is the local
compactness of the star.

Considering the values of the corresponding quantities at the edge
of the star, one can introduce their global counterparts:
$\varrho_*:=\varrho(\rho_*,\rho_C)=\varrho(\rho_*,\rho_*,\rho_C)$,
$\varsigma_*:=\varsigma(\rho_*,\rho_C)=\varsigma(\rho_*,\rho_*,\rho_C)$
and $f_*:=f(\rho_*,\rho_C)=f(\rho_*,\rho_*,\rho_C)$.

We are considering in details only the scale properties of the
solutions of ETOV system for relativistic stars. The corresponding
non-relativistic equations have the same scale properties, because
they can be considered as a special case of the relativistic ones,
taking the limit $c\to\infty$ \cite{books}.

\section{The Solution of the  ETOV System in Basic Regular Gauge}

\subsection{The General Properties of BGR Inner Solution}

The basic regular gauge (BRG) is defined by the condition
$\bar\Gamma_r\equiv 0$. It has been proved to have a unique and
important mathematical and physical properties \cite{F03}.

Together with  Eq. (\ref{Gamma_r}) the BRG definition gives a
second order differential equation for the function
$\rho_{BRG}(r)$, supplemented by the boundary conditions
\ben \rho_{BRG}(0)=\rho_C,
\,\,\,\rho_{BRG}(r_*)=\rho_*\la{BGR_bound_cond}\een
with unknown value of the radial variable $r_*$. A simple
integration of the differential equation gives:
\ben
{{\rho^\prime}\over{\rho^2}}\sqrt{{-g_{\rho\rho}}\over{g_{tt}}}=const.
\la{FirstInt}\een
After one more integration of equation $\bar\Gamma_r\equiv 0$, in
the inner domain $\rho\in [\rho_C,\rho_*]$ we obtain the relation:
\ben {{r}\over{r_*}}=\int_{\rho_C}^{\rho}{{d\rho}\over{\rho^2}}
\sqrt{{-g_{\rho\rho}}\over{g_{tt}}}\Bigg/
\int_{\rho_C}^{\rho_*}{{d\rho}\over{\rho^2}}
\sqrt{{-g_{\rho\rho}}\over{g_{tt}}}.\la{rho_r}\een
We have used the boundary condition (\ref{BGR_bound_cond}) at the
center $C$ and at the edge of the star to fix the unknown
integration constants after the integration of Eq.
(\ref{FirstInt}).

The equation (\ref{rho_r}) fixes the BRG function $\rho_{BRG}(r)$
in the interior of the star in the form
\ben\rho_{BRG}(r)=\rho_{BRG}^{int}\left({r\over{r_*}};\rho_*,\rho_C\right),
\,\,\,\hbox{for}\,\,\,r\in[0,r_*],\,\,\,\,\, \la{rho_int}\een and
yields the following basic properties of this function: \vskip
0.2truecm i)
$\rho_{BRG}^{int}\left(\eta;\rho_*,\rho_C\right)\in[\rho_C,\rho*]$
for $\eta={{r\,}\over{r_*}}\in [0,1]$;

ii) $\rho_{BRG}^{int}\left(0;\rho_*,\rho_C\right)=\rho_C$;

iii) $\rho_{BRG}^{int}\left(1;\rho_*,\rho_C\right)=\rho_*$;

iv) $\rho_{BRG}^{int}\left(\eta;\rho_0,\rho_0\right)=\rho_0$ for
$\eta\in [0,1]$.

v) $d\rho / \ d\eta\geq 0$.

vi) $\rho_C+(\rho\!_*\!-\!\rho_C)\eta\!\leq
\!\rho_{BRG}^{int}\left(\eta;\!\rho\!_*\!,\!\rho_C\right)\!\leq\!
\rho_*\!-\!(\rho\!_*\!-\!\rho_C)(1-\eta)^2$. \vskip 0.2truecm

\begin{figure}[htbp] \vspace{7.truecm}
\includegraphics{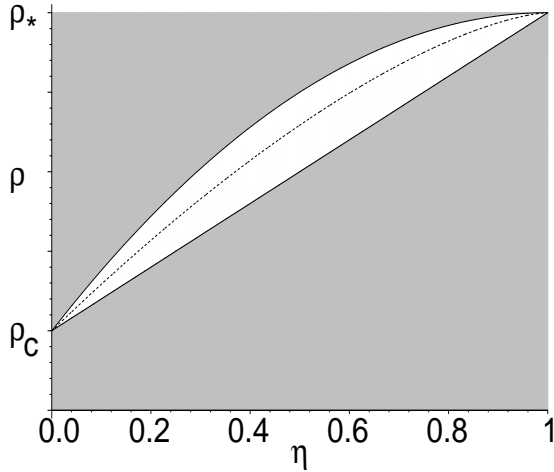} \caption{\hskip 0.2truecm The narrow constrained
physical domain of the function
$\rho=\rho_{BRG}^{int}\left(\eta;\!\rho\!_*\!,\!\rho_C\right)$,
defined by its properties  i) -- vi).
    \hskip 1truecm}
    \label{Fig0}
\end{figure}

These general properties entail the representation:
$$\rho_{BRG}^{int}\left(\eta;\!\rho\!_*\!,\!\rho_C\right)\!=\!
\rho_C\!+\!(\rho\!_*\!-\!\rho_C)\eta\!+\!(\rho\!_*\!-\!\rho_C)
\eta(1\!-\!\eta)q(\eta;\!\rho\!_*\!,\!\rho_C)$$ with some
nonnegative function $q(\eta;\rho_*,\rho_C)\in [0,1]$, which is
bounded and continuous in the 3D domain $\{\eta\in [0,1],\,
\rho_*\geq\rho_C>0\}$.

Actually the function $q(\eta;\rho_*,\rho_C)$ is to be find from
Eq. (\ref{rho_r}) and depends on the HG solution of ETOV system
with given EOS. As seen in Fig.\ref{Fig0}, this dependence is
quite weak.

Replacing the luminosity variable $\rho$ in corresponding
HG-expressions (\ref{SolBound:xyzt}) by the function
(\ref{rho_int}), in the interior of the star we obtain the stelar
quantities in BRG: \ben
m\!=\!m^{BRG}\left(\!{r\over{r_*}};\rho_*,\rho_C\!\right)\!,\,
m_0\!=\!m_{0}^{BRG}\left(\!{r\over{r_*}};\rho_*,\rho_C\!\right),
\nonumber\\
p\,=\,p^{BRG}\left(\!{r\over{r_*}};\rho_*,\rho_C\!\right),\,\,\,
\varepsilon=\varepsilon^{BRG}\left(\!{r\over{r_*}};\rho_*,\rho_C\!\right)\,,\,\,
\nonumber\\
\varphi=\varphi^{BRG}\left(\!{r\over{r_*}};\rho_*,\rho_C\!\right);
\,\,\,\hbox{-- for}\,\,\,r\in[0,r_*]. \hskip 1.4truecm
\la{BRGform}\een

\subsection{The Matching of the BGR Inner and Exterior Solutions}

The  BRG-solution in the exterior vacuum domain can be obtained
making use of Birkhoff theorem in corresponding BRG radial
variable $r$. According to \cite{F03} it is:
\ben
\rho_{BRG}^{ext}(r)\!=\!{{2m_*}\over{1\!-\!\varrho_*^2e^{2r/m_*}}}\,\,\,
\hbox{for}\,\,\,r\!\in\!\big[r_*,m_*\ln({1/{\varrho_*}})\big).
\hskip .6truecm\la{rho_ext}\een

Now we have to match interior and exterior solutions using proper
{\em physical} requirements on the luminosity variable $\rho$:

1) Because the luminosity of physical source of any kind of
radiation is $L=const/4\pi\rho(r)^2$, we see that in absence of
surface sources of the corresponding radiation the function
$\rho(r)$ must be continuous. Otherwise we will destroy the local
energy conservation at the place, where $\rho(r)$ has a jump.

2) The jumps of the derivative $\rho^\prime(r)$ will induce jumps
of the radial derivative of the luminosity, $L^\prime(r)$. This
means a jumps of the radial derivative of surface energy density.
Such phenomenon is physically possible only in presence of some
surface agent, like surface force (surface pressure) due to some
surface tension. Hence, in this case the star will have some thin
crust. Similar phenomenon is familiar, for example, from theory of
neutron stars \cite{books,ns}, where the crust is introduced and
studied from different point of view.

The presence of the crust will obviously yield an observable
consequences. Indeed, in this case, according to Eq.
(\ref{FirstInt}), the coefficient $\bar\Gamma_r$ in d'Alembert
operator $g^{\mu\nu}\nabla_\mu\nabla_\nu=g^{\mu\nu}
\left(\partial_\mu\partial_\nu-\bar\Gamma_\mu\partial_\nu\right)$
will be singular at the stelar surface:
\ben \bar\Gamma_r=-\digamma\delta(r-r_*).\la{SingBGamma}\een
Here $\delta(r)$ is 1D Dirac delta function. We refer to the
quantity $\digamma$ as to {\em "crust parameter"}. Its role in
field propagation through the stelar surface will be considered in
the next Subsection.

If we exclude the presence of stelar crust and corresponding jump
of the radial derivative $L^\prime(r)$, the crust parameter will
be zero: $\digamma=0$ and the derivative $\rho^\prime(r)$ of the
luminosity variable will be continuous function

It is obvious that the justification of the matching conditions is
impossible without right physical interpretation of the luminosity
variable $\rho$.

The above physical considerations entails the following
mathematical consequences:

1. The continuity condition
$\rho_{BRG}(r_*\!-\!0)\!=\!\rho_{BRG}(r_*\!+\!0)$ gives the basic
relation \ben
r_*=m_*\ln\left({1\over\varrho_*}\sqrt{1-{{2m_*}\over{\rho_*}}}\right)
<r_\infty. \la{r_star}\een  The {\em finite} value
$r_\infty:=m_*\ln(1/\varrho_*))$ corresponds to the physical
space-infinity in BGR, i.e., to the geometric place of points in
3D space, where the luminosity variable $\rho(r)\to\infty$ for
$r\to r_\infty-0$ and the 4D spacetime is asymptotically flat.

Hence, as in the case of point particle source of gravity
\cite{F03}, in BRG we have to consider the finite interval
$r\in[0,r_\infty)$ as a real physical domain of the radial
variable $r$. In comparison with the point particle problem, the
difference is that in stelar models we have different forms --
(\ref{rho_int}) and (\ref{rho_ext}) of the gauge function
$\rho_{BGR}(r)$ in the interior domain of the star and in the
exterior vacuum domain.

2. One can easily find that for the exterior solution
(\ref{rho_ext}) the value of the constant in Eq. (\ref{FirstInt})
is $const=1/m_*^2$.

Taking into account that:

a) The quantities $\rho$, $g_{\rho\rho}$ and $g_{tt}$ are
continuous functions of the radial variable at the point $r=r_*$;
and

b) The derivative $\rho_{BRG}^\prime(r)>0$ is positive everywhere
in the physical domain of the BGR-radial variable $r$;

\noindent we can describe the jump of this derivative by the
formula
\ben \rho_{BRG}^\prime(r_*\!+\!0)=
e^{-\digamma}\rho_{BRG}^\prime(r_*\!-\!0),\,\,\,
\digamma\in(-\infty,\infty). \hskip .6truecm\la{rho_p_jump}\een
Obviously, for $\digamma\neq 0$ this formula describes a
refraction of the lines $\rho=\rho(r)$ at the point $r_*$. It
yields the relation \ben e^{\digamma}={{m_*}^2\over{r_*}}
\int_{\rho_C}^{\rho_*}{{d\rho}\over{\rho^2}}
\sqrt{{-g_{\rho\rho}}\over{g_{tt}}}.\la{ContPrime}\een

Now, making use of the already found HG functions
$m_*(\rho_*,\rho_C)$ and $m_0(\rho_*,\rho_C)$, and matching
conditions (\ref{r_star}) and (\ref{ContPrime}), we can solve the
algebraic system of four equations
\ben
m_*(\rho_*,\rho_C)- m_*=0,\nonumber\\
m_0(\rho_*,\rho_C)- m_{0*}=0,\nonumber\\
\left({{m_*}\over{m_{0*}}}\right)^2\!\exp\!\left(2{r_*}\over{m_*}\right)\!+\!
{{2m_*}\over{\rho_*}}-1=0,\nonumber\\
{m_*}^2 \int_{\rho_C}^{\rho_*}{{d\rho}\over{\rho^2}}
\sqrt{{-g_{\rho\rho}}\over{g_{tt}}}-r_*e^{\digamma}=0\,\,\,
\la{algebraic}\een
for six unknowns $m_*,m_{0*},\rho_C,\rho_*,r_*,\digamma$ with
respect to the first four of them. Thus we arrive at a new form of
our solutions for relativistic stelar models with given EOS:
\ben m_*=m_{*}^{\,BRG}(r_*,\digamma),\,\,
m_{0*}=m_{0*}^{\,BRG}(r_*,\digamma),\nonumber\\
\rho_*=\,\rho_{*}^{\,BRG}(r_*,\digamma)\,,\,\,\,\,
\rho_{C}\,=\,\rho_{C}^{\,BRG}(r_*,\digamma).
\,\,\la{rSFParametric}\een

This representation  sheds a new light on the physical meaning of
the two-parameter family of relativistic stars, obtained in
present article:  The constant parameter $\digamma$ defines the
properties of stelar crust. For different values of this parameter
we obtain relativistic stars with different crusts.

After all, if we fix the value of the parameter $\digamma$, we
will obtain one parameter family of relativistic stars, precisely
as in Newton theory of stars and in the standard relativistic
approach to stelar physics \cite{books}, but without extra
condition $\rho_C=0$.

For example, postulating continuity of the derivative
$\rho_{BRG}^\prime(r)$ at the stelar edge $r=r_*$, we obtain
$\digamma=0$.

The existence of one parameter family of relativistic stars with
arbitrary fixed value of the parameter $\digamma$ becomes possible
just for the sake of matching conditions (\ref{r_star}) and
(\ref{ContPrime}). The condition (\ref{ContPrime}) replaces the HG
extra condition $\rho_C=0$ and produces a new type of stiff
relations in stelar physics.

It is obvious that one can impose {\em only} one of these
alternative extra conditions. A novel problem in stelar
astrophysics is to verify which one of them, if any, takes place
in Nature.

\subsection{Spreading of Waves and Static Fields Trough the Stelar Crust}

The physical agent, which brings into being the stelar crust,
changes the space-time geometry in accord to condition
(\ref{rho_p_jump}). Therefore the presence of the crust will
influence the spreading of all possible physical wave fields:
scalar, electromagnetic, gravitational, spinor, e.t.c.

In this subsection we will present a preliminary investigation of
the spreading of waves and static fields through the stelar crust.
Our aim is to reach qualitative understanding of possible role of
the stelar crust for field's dynamics and statics. Therefore we
consider in proper approximation only the simplest case of scalar
spherically symmetric field $\Phi(t,r)$. To distinguish the
effects, caused by the stelar crust, here we neglect the
interaction of the wave fields with the stelar matter. The exact
treatment of this issue is a complicated problem. Its
consideration requires first to have a complete solution for some
specific background stelar model.

As a result of Eq. (\ref{SingBGamma}) one obtains for field
$\Phi(t,r)$ in BRG the following wave equation:
\ben g^{\mu\nu}\!\nabla_\mu\!\nabla_\nu\! \Phi\!=\!
g^{tt}\Phi_{tt}\!+\!g^{rr}\Phi_{rr}\!+\!\digamma\delta(r-r_*)\Phi_r\!=\!0.\,\,\,
\la{waves}\een
Owing to the continuity of functions $g^{tt}(r)$ and $g^{rr}(r)$
at point $r_*$, we can replace them in a small enough vicinity of
the crust with their constant values $g^{tt}(r_*)$ and
$g^{rr}(r_*)$. Then, changing the corresponding scales of time and
radial variables : $t\rightarrow \sqrt{g_{tt}(r_*)}\,t$ and
$r\rightarrow \sqrt{g_{rr}(r_*)}\,r$, and using the properties of
Dirac $\delta$-function, we arrive at the differential equation:
\ben \Phi_{tt}\!-\!\Phi_{rr}\!+\!\digamma\delta(r-r_*)\Phi_{r*}=0,
\la{PDE}\een
where $\Phi_{r*}=\Phi_r(t,r_*)$ is the value of the first
derivative $\Phi_r(t,r)$ at the stelar edge.

The general solution of this equation can be represented in the
form of Fourier integral:
\ben \Phi(t,r)=\int_{-\infty}^{\infty}d\omega e^{i\omega t}
R(r;r_*,\omega). \la{PhiGen}\een
The amplitudes $R(r;r_*,\omega)$ are described by the general
solution
\ben R(r;r_*,\omega)=R(r_*,\omega)\cos\big(\omega(r-r_*)\big)+
\hskip 1.6truecm\nonumber\\
R_r(r_*,\omega)\Big(1+\digamma\Delta\Theta\big(\omega(r-r_*)\big)\Big)
{{\sin\big(\omega(r-r_*)\big)}\over\omega}\hskip 1.0truecm
\la{R_Gen}\een
of the second order ordinary differential equation:
\ben R_{rr}\!+\omega^2R\!=\!\digamma\delta(r-r_*)R_r(r_*,\omega).
\la{REq}\een
Here the arbitrary functions $R(r_*,\omega)=R(r_*;r_*,\omega)$ and
$R_r(r_*,\omega)=R_r(r_*;r_*,\omega)$ appear as integration
constants of Eq. (\ref{REq}) and for corresponding values of
$\omega$ present the values of the function $R(r;r_*,\omega)$ and
its first derivative $R_r(r;r_*,\omega)$ at the edge of the star;
$\Theta(x)$ is the Heaviside step function. For our purposes we
have to regularize this generalized function \cite{Gelfand}, i.e.,
we have to prescribe some definite value $\Theta(0)$ to this
function at the point $x=0$. Then
$\Delta\Theta(x):=\Theta(x)-\Theta(0)$.

Now it becomes clear that:

1. The physical role of the stelar crust is to produce a jump in
the $\sin$-mode of the stationary waves (\ref{R_Gen}). At the same
time the $\cos$-mode remains unchanged, crossing the crust.

2. If $\digamma=0$, i.e., in absence of stelar crust, both modes
spread trough the edge of the star as a completely free waves.

3. Choosing $R(r_*,\omega)=0$ and proper special values of
$\Theta(0)$, one can obtain solutions, which describe stationary
waves only inside the star:
$$R^{ins}(r;r_*,\omega)\!=-
\digamma R_r(r_*,\omega)\Theta\big(-\omega(r\!-\!r_*)\big)
{{\sin\big(\omega(r\!-\!r_*)\big)}\over\omega},$$ --  for
$\Theta(0)=1+1/\digamma$, or only outside the star:
$$R^{outs}(r;r_*,\omega)\!=\!
\digamma R_r(r_*,\omega)\Theta\big(\omega(r\!-\!r_*)\big)
{{\sin\big(\omega(r\!-\!r_*)\big)}\over\omega}$$ --  for
$\Theta(0)=1/\digamma$.

Thus we see that the parameter $\Theta(0)$ plays the role of
reflection coefficient for the $\sin$-mode. For proper values of
this coefficient we have a total inner, or total outer reflection
of the $\sin$-mode by the stelar crust.

4. In the static limit $\omega\to 0$ one obtains from Eq.
(\ref{R_Gen}):
\ben \Phi(r)\!=\!\Phi(r_*)\!+\!
\Phi_r(r_*)\big(1\!+\!\digamma\Delta\Theta(r\!-\!r_*)\big)(r\!-\!r_*).
\hskip .5truecm \la{StaticPhi}\een
The last formula shows that crossing the stelar crust with
$\digamma \neq 0$, the static field $\Phi(r)$ is a subject of
refraction.

The existence of the stelar crust with the above properties is a
new specific prediction of our models of relativistic stars. It
may have important consequences not only for the stelar physics
and needs further careful study.

\section{The Solution of the  ETOV System in Physical Regular
Gauge} The regular change of the rho-gauge, defined by the
fractional linear mapping of the interval $r\in [0,r_\infty)$ onto
the whole interval $r\in[0,\infty)$: \ben r \to
m_*\ln(1/\varrho_*){{r}\over{r+m_*/\ln(1/\varrho_*)
}}\la{rrgt}\een brings us to the physical regular gauge (PRG)
\cite{F03}. There the radial variable $r$ varies in the standard
semi-infinite interval $[0,r_\infty)$. (Note that we are using the
same notations $r,\,r_*,...$ for radial variables in different
$\rho$-gauges.) Then in PRG we have \ben
\rho_{PRG}(r)=\begin{cases}
\rho_{BRG}^{int}\left(\eta;\rho_*,\rho_C\right)\,\,\,\,,
&\text{if $r\in [0,r_*]$} ;\\
2m_*\left(1-e^{2\varphi_G}\right)^{-1}, & \text{if $r\in
[r_*,\infty),$}
\end{cases}\hskip .6truecm\la{rho_PRG} \een
where now \ben
\eta={{r(r_*+m_*/\ln(1/\varrho_*))}\over{r_*(r+m_*/\ln(1/\varrho_*))}}
\in[0,1]\,\,\,\hbox{for}\,\,\,r\in[0,r_*],\hskip .6truecm
\la{eta_PRG}\een \ben r_*\!=\!m_*\left(\!{1/{\ln\left({1}\over
{\sqrt{1\!-\!{{2m_*}\over{\rho_*}}}}\right)}}\!-\!
1/\ln\left({1\over\varrho_*}\right)\!\right)\geq 0.\hskip .6truecm
\la{r_star_PRG}\een In Eq. (\ref{rho_PRG}) we are using the
modified Newton gravitational potential \cite{F03}: \ben
\varphi_{\!{}_G}(r;m_*,m_{0*}):=-{{m_*}\over
{r+m_*}/\ln({m_{0*}\over m_*})}. \la{Gpot} \een

From Eq. (\ref{rho_PRG}) one easily obtains the important
inequality for the stelar parameters, written in the following two
useful forms: \ben f_*:=\varrho_*^2+{{2m_*}\over{\rho_*}}\leq
1\,\,\,\Leftrightarrow\,\,\,\rho_*\geq
{{2m_*}\over{1-\varrho_*^2}}. \la{f<1}\een This is a more strong
restriction on the domain $\mathbb{D}^{(2)}_{\rho_*,\rho_C}$ than
the inequality (\ref{Drhorho:y}). Actually the inequality
(\ref{f<1}) is a direct consequence of matching condition
$\rho(r_*-0)=\rho(r_*+0)$, written in the following BRG-form:
$$ 1=\varrho_*^2e^{2{r_*}\over{m_*}}+{{2m_*}\over{\rho_*}}\Rightarrow
1\geq \varrho_*^2+{{2m_*}\over{\rho_*}}\,\,\,\hbox{for}\,\,\,
r_*\geq 0.$$

\section{Some Concluding Remarks}

It is clear that the new approach to stelar structure, developed
in the present article, calls for revision of many of widely
accepted features of the GR theory of stars. The changes are not
based on the critics of the very GR, but on more deep
understanding of its applications, and on solution of some open
problems in this theory, like the physical justification of the
choice of GR gauges.

In particular, it is obvious that we must apply the Birkhoff
theorem in PRG only in the interval $\rho\in[\rho_*,\infty)
\rightleftarrows r\in[r_*,\infty)$, i.e. in the exterior vacuum
domain outside the star. As a result, in this domain all {\em
local} GR effects like gravitational redshift, perihelion shift,
deflection of light rays, time-delay of signals, etc., are gauge
invariant and will have their standard {\em exact} values. The PRG
metric for this domain can be found in \cite{F03}.  The
differences between predictions of our general models of stars and
the standard ones, based on the assumption $\rho_C=0$, can not be
observed in the local gauge invariant {\em gravitational}
phenomena, which take place in the outer vacuum domain,
surrounding the stars.

Hence, our most general geometrical models have an essential
impact only on the theory of the interior of relativistic stars,
and on theory of spreading of different physical fields in stars,
and around the stars.

The obtained new results seems to deserve further study and can
lead to serious changes of our understanding of physics of stars
in Nature. Specific models of the described novel type for
relativistic stars with different EOS, as well as other
developments and applications to problems of real stelar physics,
will be published in subsequent articles of this series.

\vskip 1.truecm

{\em \bf Acknowledgments} \vskip .3truecm

The author is grateful to the High Energy Physics Division, ICTP,
Trieste, for the hospitality and for the nice working conditions
during his visit in the autumn of 2003. There an essential basic
ideas of present article were developed.

The author is vary grateful, too, to the JINR, Dubna, for the
financial support of the present article and for the hospitality
and good working conditions during his two three-months visits in
2003 and in 2004, when the most of the work has been done.

The author is deeply indebted to Prof. T.~L.~Boyadjiev for
friendly support and many useful discussions on mathematical
problems, related to the stelar boundary problem, to Prof.
S.~Bonazzola, for discussions, and especially, for encouraging
general information about the existence of results in direction of
overcoming of the standard relativistic restriction on the stelar
masses. Similar results can  be found in the articles \cite{Bell},
too. The author wishes to thank Prof. Ll.~Bell for attracting his
attention to these articles.

The author is tankful, too, to Prof. V.~Nesterenko and to unknown
referee of the first of the articles \cite{F03}, who raised the
problem of point particle limit of bodies of finite dimension in
GR, thus stimulating the development of general geometrical models
of relativistic stars.

\end{document}